\documentstyle[twocolumn,aps,epsf]{revtex}

\begin{document}
\draft

\twocolumn[\hsize\textwidth\columnwidth\hsize\csname %
 @twocolumnfalse\endcsname

\title{
Block Spins for Partial Differential Equations\footnote{To appear in a
special issue of {\it Journal of Statistical Physics\/} dedicated to Leo
Kadanoff on the occasion of his 60th birthday.}
}

\author{Nigel Goldenfeld$^1$, Alan McKane$^2$ and Qing Hou$^1$}

\address{
$^1$Department of Physics,
University of Illinois at Urbana-Champaign, 1110 West Green Street,
Urbana, IL. 61801-3080, USA.
}

\address{
$^2$Department of Theoretical Physics, University of Manchester,
Manchester M13 9PL, England.
}

\maketitle
\begin{abstract}
We investigate the use of renormalisation group methods to solve partial
differential equations (PDEs) numerically. Our approach focuses on {\it
coarse-graining\/} the underlying continuum process as opposed to the
conventional numerical analysis method of {\it sampling\/} it. We
calculate exactly the coarse-grained or `perfect' Laplacian operator and
investigate the numerical effectiveness of the technique on a series of
$1+1$-dimensional PDEs with varying levels of smoothness in the
dynamics: the diffusion equation, the time-dependent Ginzburg-Landau
equation, the Swift-Hohenberg equation and the damped
Kuramoto-Sivashinsky equation.  We find that the renormalisation group
is superior to conventional sampling-based discretisations in
representing faithfully the dynamics with a large grid spacing,
introducing no detectable lattice artifacts as long as there is a
natural ultra-violet cut off in the problem.  We discuss limitations and
open problems of this approach.

\end{abstract}
\vspace{0.3in}

]

\section{Introduction}

It is a rare event in science that a single paper contains an idea so
powerful that it revolutionises an entire field.   Rarer still are those
gems which transform two or more apparently separate fields, such as
the paper entitled {\it Scaling Laws for Ising Models near
$T_c$\/} by Leo P.  Kadanoff\cite{lpk}, published in the regrettably
short-lived journal {\it Physics}.  Indeed, this organ, edited and
founded by P.W. Anderson and the late B.T. Matthias, announced as its
by-line the memorable claim ``An international journal for selected
articles which deserve the special attention of physicists in all
fields"; and perhaps no other journal before or since has lived up to
this hubris.  Kadanoff's famous article developed the notion of what
came to be called ``block spins", and was cited in the title of the
early seminal paper by K.G. Wilson which introduced the modern form of
the renormalisation group (RG) in both condensed matter and high energy
physics\cite{kgw}.

Although these theoretical developments have become part of the canon of
modern physics, the ``spin-off" from Kadanoff's work continues to this
day, as mathematicians and physicists, including Kadanoff and colleagues
at the University of Chicago, study the singularities and patterns which
arise in extended physical systems, governed only by sets of partial
differential equations (PDEs)\cite{ch}.  On the auspicious occasion of
this 60th birthday {\it Festschrift\/} to honour Leo P. Kadanoff, it
therefore seems appropriate to contribute a brief account of our recent
unpublished work\cite{gm} which extends the ``block spin" insights and
renormalisation group theory to the numerical solution of PDEs.  Some,
but not all of our results have recently been rederived independently
by Katz and Wiese\cite{kw}, and we shall comment on the differences in
the sequel.  This work is part of our program to utilise RG methods for
PDEs\cite{goldenfeldbk,cgo}, and is distinct from our earlier work
applying RG iterative methods to construct similarity solutions and
travelling waves\cite{cheng}.

\section{Theory of Perfect Operators}

\subsection{Motivation}

Discretisation is an inevitable part of numerical analysis.  Let us
suppose that we wish to solve a partial differential equation
numerically.  The standard procedure in real space is to suppose that
the solution $u(x)$ is sampled at points $x_i$ and an algorithm devised
to approximate the values $u_i \equiv u(x_i)$.  If the points are
equidistant with spacing $dx$, then we naturally require that in the
continuum limit $dx\rightarrow 0$, the sequence $u_i$ converges to
$u(x)$.

The disadvantage of the sampling approach is that one is forced to
reproduce as faithfully as possible all the detail and fine structure of
the solution, even on a scale that may be of no interest or worse,
beyond the regime of applicability of the differential equation itself.
This has two consequences:

\begin{itemize}
\item a small grid size $dx$ must be used, which implies many grid points
must be calculated and stored;
\item a small time step $\delta t$ is implied by the small $dx$, either
for reasons of accuracy or stability of the numerical method.
\end{itemize}

As a result, the numerical method is subjected to an unnaturally large
degree of computational complexity.  Finally, for a given sampling
procedure there is no unique prescription for obtaining the equation
governing the sampled points $u_i$.  The only criterion for
admissability is that it converges in the continuum limit; in practice
of course, one seeks schemes which are numerically stable and attain the
continuum limit rapidly.

From the statistical physicist's point of view, Kadanoff's 1966 paper
suggests that a more natural approach to discarding information is to
{\it coarse-grain\/} rather than to {\it sample\/}.  Such a procedure
focuses on the scale of interest, whilst allowing the possibility of
accessing the continuum function to an arbitrary level of detail if
desired.  Suppose that we denote the coarse-graining operator at scale
$\Lambda$ by the symbol $C_\Lambda$, with capital letters denoting
coarse-grained quantities.  Then conceptually we need to find the
operator $L_\Lambda$ which connects $U(X,0)$ and $U(X,t)$ given the
microscopic time evolution operator $L$ connecting $u(x,0)$ with
$u(x,t)$, as shown schematically in the commutativity diagram below:

\begin{eqnarray*}
u(x,0) &\quad\mathop{\longrightarrow}\limits^{L}\quad &u(x,t)\\
C_\Lambda\Bigg\downarrow &&C_\Lambda\Bigg\downarrow\\
U_\Lambda(X,0) &\quad \mathop{\longrightarrow}\limits^{L_\Lambda}\quad & U_\Lambda(X,t)
\end{eqnarray*}

In practice, we coarse-grain onto a lattice $X_i$ with a specified
$C_\Lambda$. The freedom of choice in this coarse-graining operator
parallels the lack of uniqueness in defining a governing equation for
sampled points $u_i$ in the sampling approach; but once a
coarse-graining operator $C_\Lambda$ has been defined, there should be a
unique prescription to obtain $L_\Lambda$.

Our diagram suggests that coarse-graining and time evolution
commute, but it is not clear that this is correct, even in principle.
For example, an equation with a positive Liapunov exponent might have
the following property: two initial values $u_{1,2}(x,0)$ differing only
in field configuration at two nearby points in space, but with the same
coarse-grained representation $U(X,0)$, might differ substantially at very
long times $t$: $u_1(x,t) \neq u_2(x,t)$ even though the coarse-grained
initial fields would evolve identically.  This example raises the
interesting question of whether such equations are well-defined: no
numerical procedure would be appropriate, unless the divergence of the
trajectories was still bounded (as in a strange attractor).

We also need to consider the appropriate coarse-graining scale.  Two
situations are possible here.  In the first, we suppose that the
solution we wish to obtain has a natural scale $\Lambda$ below which
there is no significant structure.  In that case, our goal is to avoid
having to over-discretise the problem merely in order to attain the
accuracy of the continuum limit.  Thus, we would like to be able to use
as large a value for the grid spacing $dx$ as possible without
sacrificing accuracy.  In the second situation, there is no such obvious
scale, or at least, it is not known {\it a priori}, but the
computational demands are so large that it is simply not feasible to
work with a grid spacing $dx$ smaller than some size $\Lambda$.  In this
case, we would like to minimise in some sense the artifacts that must
inevitably arise.

We will mainly have in mind the first situation, which is more
straightforward because the only issue is speed of convergence to the
continuum limit: there is no explicit discarding of important dynamical
information.  Instead the focus is how to remove lattice discretisation
artifacts\cite{gm,kw}.  We will refer to an operator or equation as
being `perfect' if it has been constructed by coarse-graining
appropriately so that it has no lattice artifacts; our discussion
follows the pioneering work of Hasenfratz and Niedermeyer in the context
of lattice gauge theory\cite{hn}.

In the second situation, however, one is making an uncontrolled and
potentially severe truncation of the correct dynamics, and issues of
modelling must be faced.  For example, can one model the neglected
unresolved scales as effective renormalisations of the coefficients in
the original PDE? Are the neglected degrees of freedom usefully thought
of as noise for the retained large-scale degrees of freedom?  And how
can any available statistical information on the small-scale degrees of
freedom be used to improve the numerical solution for the large-scale
degrees of freedom?  Such questions may perhaps be treated by combining
constrained Monte Carlo simulation of the microscopic degrees of
freeedom with a maximum entropy criterion for discretisation of the
large-scale degrees of freedom.  An alternative but related approach has
been implemented by Kast and Chorin\cite{kc}, who minimise the RMS
error, estimated from knowledge of the microscopic probability
distribution.

\subsection{The Perfect Laplacian}

Our goal in this section is to examine the simplest possible problem,
namely the diffusion equation in $d + 1$ dimensions:

\begin{equation}
\partial_t u({\bf x},t) = \partial^2_x u({\bf x},t)
\label{eqn:diff}
\end{equation}  
subject to appropriate initial conditions and boundary conditions.

The coarse-graining is defined with respect to a $d$-dimensional
hypercubic lattice of spacing $a$ on whose vertices reside the lattice 
variables $U({\bf n},t)$ defined by

\begin{equation}
U({\bf n},t) = \frac{1}{a^d}\int_{-a/2}^{a/2}d^dx\, U({\bf x}+{\bf n}a,t).
\end{equation}

The numerical scheme that we will adopt is Euler discretisation in time, but
with a perfect discretisation for the Laplacian:

\begin{equation}
U({\bf n},t+dt) = U({\bf n},t) + dt \, \Delta_a U({\bf n},t)
\end{equation}
where the perfect Laplacian $\Delta_a$ at scale $a$ is given by

\begin{eqnarray}
&&e^{-\frac{1}{2}\sum_{\bf n} U({\bf n},t) \Delta_a U({\bf n},t)} = \int Du\, e^{-\frac{1}{2}\int d^dx \, u(x)\partial_x^2 u(x)}\nonumber\\
&&\qquad\times\prod_{\bf n}\delta\left(U({\bf n},t) -  \frac{1}{a^d}
\int_{-a/2}^{a/2}d^dx\, u({\bf x}+{\bf n}a,t)\right)
\label{eqn:coarse}
\end{eqnarray}
where the delta function enforces the definition of the coarse-grained field
\begin{equation}
U({\bf n},t) =  \frac{1}{a^d}
\int_{-a/2}^{a/2}d^dx\, u({\bf x}+{\bf n}a,t).
\label{eqn:coarsegrainvar}
\end{equation}
This definition of the coarse-grained operator is of course motivated by
field theory, but is unsatisfactory for several reasons.  First, it is
admittedly {\it ad hoc}; in some sense, it is imposing a probability
distribution on the high wavenumber degrees of freedom which is not
necessarily present in the actual solution of the PDE being solved.
Second, we are implicitly assuming that the coarse-graining of a
differential equation is accomplished by simply coarse-graining the
separate terms in the equation according to the prescription given
above.  We would prefer to able to start with the governing PDE and
coarse-grain the entire equation in a systematic procedure.  We hope to
report on this methodology in a future publication. The delta-function
constraint in Equation (\ref{eqn:coarse}) can be softened by writing
$\delta(x) \rightarrow C(\kappa)\exp(-2\kappa x^2)$, with $C(\kappa)$
being the normalisation.

The calculation of the functional integral in Equation (\ref{eqn:coarse}) is
straightforward\cite{hn} and yields a convolution expression

\begin{equation}
\Delta_a U({\bf n}) = -\frac{1}{a^2}\sum_{\bf R} U(({\bf n}+{\bf R})a)\rho({\bf R})
\label{eqn:perflap}
\end{equation}
where ${\bf R}$ is a $d$-dimensional lattice integer, and
\begin{equation}
\rho({\bf R}) = \int \frac{d^d{\bf p}}{(2\pi)^d}\,e^{-i{\bf p}\cdot{\bf R}} f({\bf p})
\end{equation}
\begin{equation}
f({\bf p})^{-1} =\sum_{{\bf l}=-\infty}^\infty \frac{1}{({\bf p}+2\pi {\bf l})^2}
\prod_{i=1}^d\frac{4 \sin^2 p_i/2}{(p_i + 2\pi l_i)^2} + \frac{1}{3\kappa}.
\end{equation}
Here, the vector ${\bf p}$ lies in the first Brillouin zone: $|p_i| <
\pi$.  Evaluation of the coefficients $\rho({\bf R})$ requires numerical
integration in general, but in the case of $d=1$ there is a
simplification because
\begin{equation}
f({ p})^{-1} =\sum_{{ l}=-\infty}^\infty \frac{1}{({p}+2\pi {l})^4} 4 \sin^2 p/2 + 
\frac{1}{3\kappa}
\end{equation}
which can be summed by contour integration to yield
\begin{equation}
f({ p})^{-1} = \frac{1}{4}\hbox{cosec}^2 p/2 + \left(\frac{1}{3\kappa}-\frac{1}{6}\right).
\end{equation}

\noindent
In the case $\kappa=2$, 
\begin{equation}
\rho(R)=2\delta_{R,0} - \delta_{R,1} - \delta_{R,-1}
\label{eqn:kappatwo}
\end{equation}
which looks just like the conventional lowest order finite difference
expression for the second derivative.  This is, however, slightly
misleading, because this kernel acts on the {\it coarse-grained\/}
function $U(n)$, not the sampled value of the continuum function.

The general result for arbitrary $\kappa$ is
\begin{equation}
\rho(R) = \left\{
\begin{array}{ll}
A\left(\left(\frac{3\kappa}{\kappa+4}\right)^{1/2} - 1\right) 
&\quad\hbox{if $R=0$} \\
A\left(\frac{3\kappa}{\kappa+4}\right)^{1/2} \left(\frac{(3\kappa(\kappa+4))^{1/2}-2(1+
\kappa)}{\kappa-2}\right)^{|R|} & \quad\hbox{otherwise}
\end{array}
\right.
\end{equation}

\noindent
where $A\equiv{-6\kappa}/{(\kappa-2)}$.

These formula are problematic to interpret for the cases where $\kappa <
\infty$, because the functional integral in Equation (\ref{eqn:coarse})
does not exactly enforce the definition of the coarse-grained field
$U$.  Thus, for example, is it consistent to coarse-grain the initial
condition using Equation (\ref{eqn:coarsegrainvar}), whilst at the same
time using the perfect Laplacian operator with $\kappa <
\infty$?  We will return to this issue below.

\section{Numerical Results in One Dimension}

\begin{figure}[h]
\begin{center}
\leavevmode
\hbox{\epsfxsize=0.9\columnwidth \epsfbox{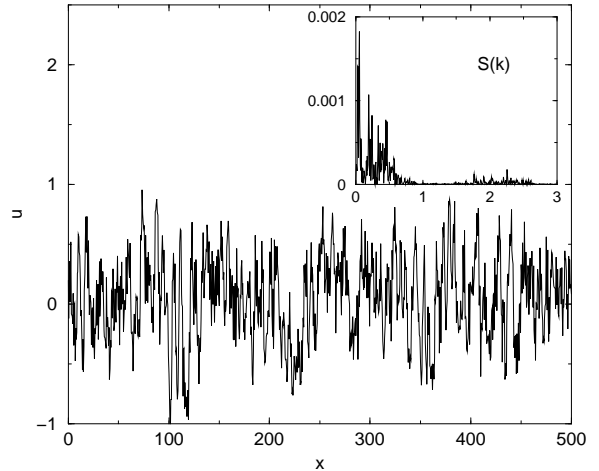}}
\end{center}
\caption{Random initial condition with $dx=0.5$, $N=1024$ grid points.
The inset shows the power spectrum of the initial condition, $S(k)$.}
\label{fig:fig1}
\end{figure}

In this section, we present numerical results obtained on differential
equations in one space and one time dimension, but with varying
character of solutions.  All the results were initiated from the random
initial conditions shown in Figure (\ref{fig:fig1}), defined on a
lattice of 1024 points with grid spacing $dx=0.5$ with periodic boundary 
conditions.

First we examine the diffusion equation, where
high wavenumber behaviour dies away rapidly.  As expected, the RG
numerical method performs well compared to the standard algorithms.  The
second example is the coarsening dynamics of the time dependent
Ginzburg-Landau equation (sometimes known as model A), where the
coalescence of domains is the dominant behaviour; we will see that
over-aggressive coarse-graining of the initial conditions can lead to
late time configurations which differ from the correct solution.
Although it seems plausible that the statistical information is
preserved, i.e. the ensemble averaged structure factor, we have not yet
explicitly checked this hypothesis.  The third example is the
Swift-Hohenberg equation, which in the parameter range we studied forms
a lamellar phase with a well-defined periodicity.  Again,
over-aggressive coarse-graining is seen to be counter-productive.
Lastly, we studied the damped Kuramoto-Sivashinsky equation, which is a
toy model for directional solidification and other systems which form
interfacial patterns.  The interesting aspect of this benchmark is that
a spatio-temporal chaotic phase exists for certain parameter ranges.  As
one might expect from the heuristic comments earlier, faithful
reproduction of the solution is not really possible with any significant
degree of coarse-graining.

In order to perform these numerical experiments, we have had to make two
uncontrolled approximations.  First, we have followed Katz and Wiese and
used the exact coarse-graining operation Equation
(\ref{eqn:coarsegrainvar}) but allowed the Laplacian operator to have
any $\kappa$, not just the value $\kappa=\infty$.  In particular, we
have used the special value $\kappa = 2$, where the Laplacian takes on
its conventional and very local form.  Our results are essentially
unchanged when we use the $\kappa=\infty$ form of the Laplacian, but
there is a slight loss of stability of the Euler algorithm in this case,
which sets a limit on the maximum value of $dt/dx^2$.  For a general
value of $\kappa$, the stability limit (calculated for the diffusion
equation) is given by 
\begin{equation}
\frac{dt}{dx^2} < \frac{1}{6} + \frac{2}{3\kappa}, 
\end{equation} 
implying that $dt/dx^2 < 1/6$ for $\kappa = \infty$.

Second, we have not yet calculated the coarse-grained operators
corresponding to nonlinear operators such as $u^3$ and $u\partial_xu$:
instead we have simply replaced each nonlinear operator $N(u)$ by
writing $C_\Lambda N(u) = N(U)$.  This approximation can be controlled
in a more systematic derivation of the theory.

\subsection{Diffusion Equation}

We used the random initial condition shown in Figure (\ref{fig:fig1}),
defined on a lattice of 1024 points with grid spacing $dx=0.5$, and
evolved it in three different ways to time $t=100$.  The first way was a
benchmark calculation using a conventional numerical analysis Euler
scheme with $dt=0.001$.  

\begin{figure}[h]
\begin{center}
\leavevmode
\hbox{\epsfxsize=0.9\columnwidth \epsfbox{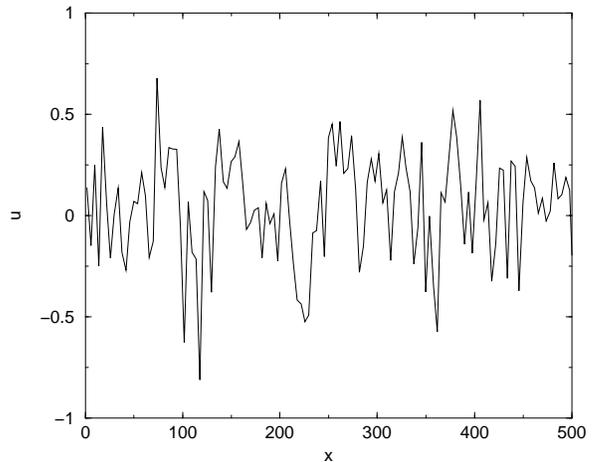}}
\end{center}
\caption{Random initial condition coarse-grained so that $dx=4$ 
and $N=128$ grid points.}
\label{fig:fig2}
\end{figure}

The second way used the RG methodology we have described above.  We
coarse-grained the initial condition down to a smaller lattice of 128
grid points and $dx=4$, using the coarse-graining of Equation
(\ref{eqn:coarsegrainvar}), yielding the function exhibited in Figure
(\ref{fig:fig2}).  We evolved this forward in time using the perfect
Laplacian of Equation (\ref{eqn:kappatwo}) with $\kappa=2$, and $dt=5$.
Such a large value was stable because of the much larger value of $dx$
than in the benchmark.

In the third way, we sampled $N=128$ points of the initial condition 
with a uniform grid spacing of $dx=4$, and then evolved this forward in time
using the standard Euler method.  

The appropriate comparison is between the two calculations using 128 points:
how well does each reproduce the benchmark calculation?

\begin{figure}[h]
\begin{center}
\leavevmode
\hbox{\epsfxsize=0.9\columnwidth \epsfbox{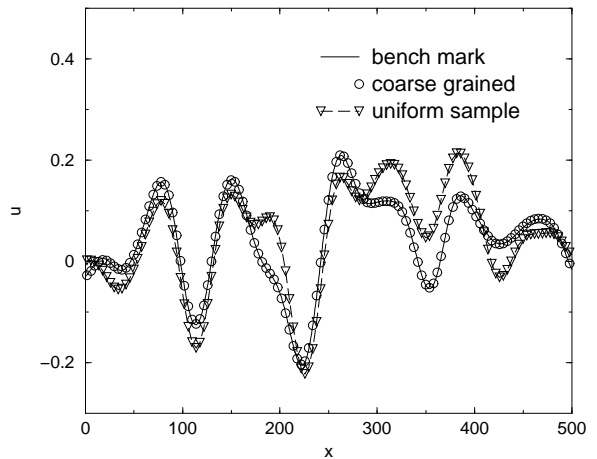}}
\end{center}
\caption{Comparison of results for the diffusion equation for $t=100$
and $N=128$.  The solid line represents the benchmark, the circles the
coarse-grained method, and the triangles the conventional
sampling method.}
\label{fig:fig3}
\end{figure}

In Figure (\ref{fig:fig3}) is plotted the benchmark configuration at
$t=100$ along with the results from the coarse graining or sampling
methods. The coarse grained result almost exactly falls onto the
benchmark calculation, even though it uses an 8-fold coarser lattice and
a bigger time step size. Numerical stability dictates that $dt <
dx^2/2$. This means that the largest time step size one can use for the
benchmark lattice is about $dt=0.1$.  Therefore, by coarse graining by a
factor of eight and using a $dt$ 50 times larger than the maximum
permitted for the benchmark, the calculation was accelerated by a factor
of 400, without introducing any noticeable lattice artifacts. The
reduction in CPU time will be more pronounced when we go to higher
dimensions.  Furthermore, one sees clearly that with the same degree of
discretisation, the uniform sampling result is inferior to the
coarse-graining result and does not reproduce the benchmark calculation
very well.

The reason that coarse-graining works well in this instance is because
our procedure preserved all relevant information down to the coarse
grained scale $\Lambda$. In the uniform sampling method, however, one
grid point is used to represent all those within a neighbourhood of size
$\Lambda$; thus, this point is likely to be in the tail of the spectrum
spanned by function values in the neighbourhood.

What is an appropriate scale at which to coarse-grain?  $\Lambda$ should
be set such that the time scale of interest is larger than the
relaxation time $\tau_{relax}$ of details on a scale smaller than
$\Lambda$. In the case of the diffusion equation, if the
initial configuration is very smooth on the scale of $\Lambda$, we can
safely coarse grain to that level; the inset to Figure (\ref{fig:fig1})
shows the power spectrum of the initial data, and reveals that the
coarse-graining level used preserves the salient long-wavelength
features. When there are significant high wavenumber modes, given that
$\tau_{relax}$ of a fluctuation on the order of $dx$ is roughly
proportional to $dx^2$, we can at most coarse grain to a level $\Lambda
\sim \sqrt{T}$ with $T$ being the time scale of interest.  Because
computation is most consuming in simulations with long evolution time and
large lattices, we almost always want to coarse grain to some level.
But if we are interested in the early time evolution of the
configuration with significant high wavenumber modes, we should not
coarse grain at all.

\subsection{Time-dependent Ginzburg-Landau Equation}

The time-dependent Ginzburg-Landau equation was studied with the discretisation

\begin{equation}
u(n,t+dt) = u(n,t) + dt \,[ \epsilon \, u(n,t) - u(n,t)^3  + \Delta_a u(n,t)].
\end{equation}

\begin{figure}[h]
\begin{center}
\leavevmode
\hbox{\epsfxsize=0.9\columnwidth \epsfbox{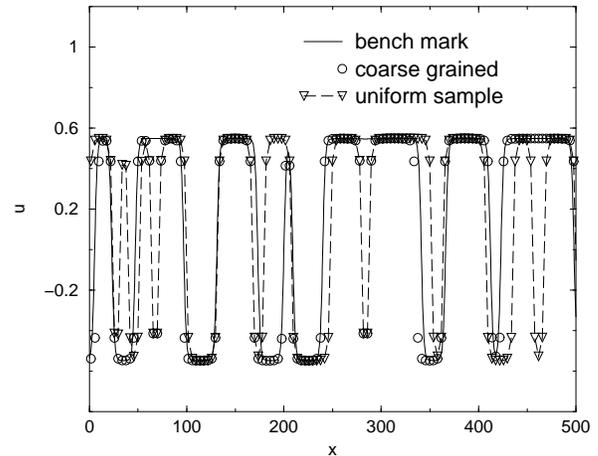}}
\end{center}
\caption{Comparison of results for the TDGL equation with $\epsilon=0.3$
for $t=100$ and
$N=128$.  The solid line represents the benchmark, the circles the
coarse-grained method, and the triangles the conventional
sampling method.}
\label{fig:tdgl8}
\end{figure}

It has a nontrivial fixed point $u(n,t) = \pm\sqrt{\epsilon}$ when
$\epsilon$ is positive. Unlike the diffusion equation, where the
configuration continually flattens, one sees
regions of $u(n,t) = \sqrt{\epsilon}$ and $u(n,t) = -\sqrt{\epsilon}$
forming, separated by domain walls. As time progress, these regions
coalesce and expand as shown in Figure (\ref{fig:tdgl8}).

In our simulation, $\epsilon=0.3$, and the benchmark lattice was evolved
with $dx=0.5$, $dt=0.001$. Coarse graining and sampling were used to
evolve the system with $dx=2$, $dt=0.1$, and $dx=4$, $dt=0.001$. As
shown in Figure (\ref{fig:tdgl8}), coarse graining always yields
superior results than uniform sampling. Also, one notices that when
coarse grained to $dx=4$, there are some suprious offshoots in the
configuration. This is because we coarse-grained too aggressively.
The structure factor $S(k)\equiv|u(k)|^2$ of the initial configuration
is plotted in Figure (\ref{fig:fig1}). It has significant spectral
weight beyond $k=\pi/8$.  Each coarse-graining step (defined as double
the $dx$ size) discards half of the Brillouin zone with larger
wavenumber. So coarse-graining three times loses the modes with $k >
\pi/8$. This is not serious for diffusion where the dynamics is a
trivial decay of large wavenumber modes, but is a poor approximation for
model A where domains grow from the initial configuration.

\begin{figure}[h]
\begin{center}
\leavevmode
\hbox{\epsfxsize=0.9\columnwidth \epsfbox{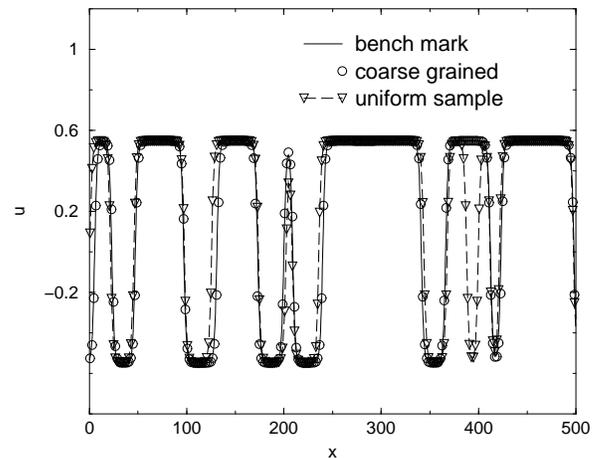}}
\end{center}
\caption{Same as Figure (4) but with $N=256$.}
\label{fig:tdgl4}
\end{figure}

In Figure (\ref{fig:tdgl4}) we have compared the results when a smaller 
$dx$ is used: with this
resolution, the coarse-graining algorithm still captures most of the
correct features of the solution, whereas the sampling method
introduces a spurious domain. 

\subsection{Swift-Hohenberg Equation}

\begin{figure}[h]
\begin{center}
\leavevmode
\hbox{\epsfxsize=0.9\columnwidth \epsfbox{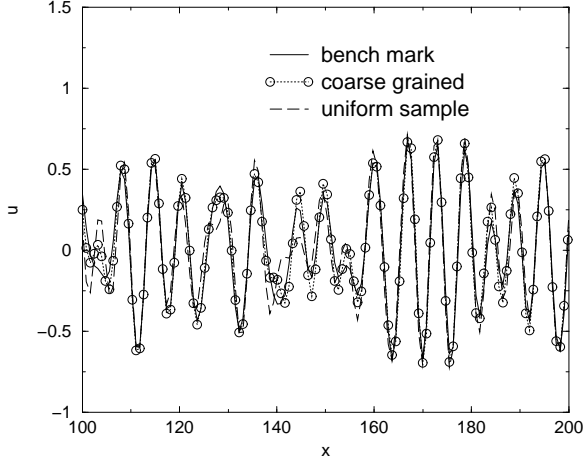}}
\end{center}
\caption{Solution of the Swift-Hohenberg equation with $\epsilon=0.5$,
$dx=\pi/4$, $dt=0.001$, $t=4$. The solid line represents the benchmark, the circles the
coarse-grained method, and the triangles the conventional
sampling method.}
\label{fig:hs1}
\end{figure}

The discretised form of the Swift-Hohenberg equation\cite{sh}

\begin{eqnarray}
u(n,t+dt) = u(n,t) &&+ dt \,[ \epsilon \, u(n,t) - u(n,t)^3\nonumber\\
 &&- (1+\Delta_a)^2 u(n,t)]
\end{eqnarray}

\noindent
was studied using the comparison methodology described above, and the
results are shown in Figure (\ref{fig:hs1}).

\begin{figure}[h]
\begin{center}
\leavevmode
\hbox{\epsfxsize=0.9\columnwidth \epsfbox{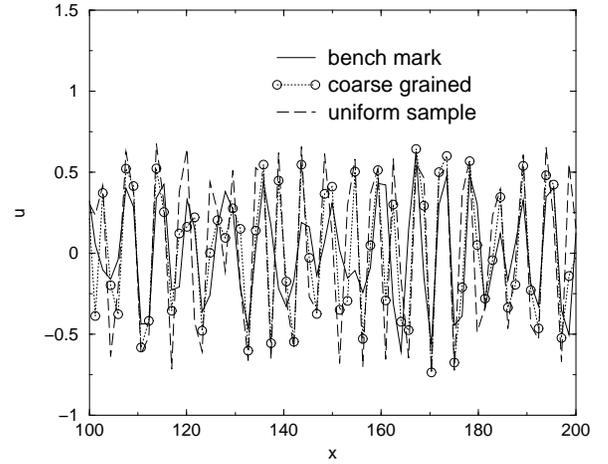}}
\end{center}
\caption{Solution of the Swift-Hohenberg equation with $\epsilon=0.5$,
$dx=\pi/2$, $dt=0.001$, $t=4$. The solid line represents the benchmark, the circles the
coarse-grained method, and the triangles the conventional
sampling method.}
\label{fig:hs2}
\end{figure}

Because of the $(1+\Delta_a)^2$ term, there is an intrinsic length scale
of $2\pi$ and a periodic ground state. The dynamics is interesting
because $u$ not only has an amplitude but also a phase. We choose
$\epsilon=0.5$. The benchmark lattice was evolved with $dx=\pi/16$,
$dt=0.0001$ to time $t=4$. The coarse-graining and sampling methods used
$dx=\pi/4$, $dt=0.001$. Again, the coarse-grained result reproduces the
benchmark quite well, better than sampling, but not as well as found in
experiments on the diffusion and TDGL equations. We attribute this to
the existence of the intrinsic length scale. When we coarse-grain to a
level close to this period, many relevant modes are ignored. Indeed,
when we coarse-grain further to $dx=\pi/2$, the dynamics longer follows the
benchmark very closely, as shown in Figure (\ref{fig:hs2}).

\subsection{Damped Kuramoto-Sivashinsky Equation}

We also considered the damped Kuramoto-Sivashinsky equation\cite{dks},
which can undergo a phase transition from a lamellar phase to
spatio-temporal chaos when the control parameter $\epsilon$ exceeds a
critical value $\epsilon_c\sim 0.68$. The discretised equation of motion studied was

\begin{eqnarray}
u(n,t+dt) = u(n,t) + &&dt \,[ \epsilon \, u(n,t) - u(n,t) \nabla u(n,t) \nonumber\\
&&\quad - (1+\Delta_a)^2 u(n,t)]
\end{eqnarray}

\noindent
where $\nabla u(n,t)$ is the usual central difference formula for first derivative.


\begin{figure}[h]
\begin{center}
\leavevmode
\hbox{\epsfxsize=0.9\columnwidth \epsfbox{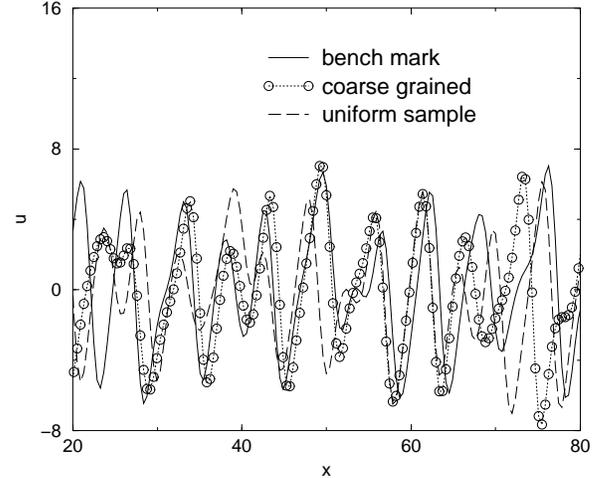}}
\end{center}
\caption{Solution of the damped Kuramoto-Sivashinsky equation with $\epsilon=0.9$,
$dx=\pi/8$, $dt=0.0001$, $t=16$. The solid line represents the benchmark, the circles the
coarse-grained method, and the triangles the conventional
sampling method.}
\label{fig:dks1}
\end{figure}

We examined both the lamellar phase ($\epsilon=0.4$) and the chaotic phase
($\epsilon=0.9$). The former situation is very similar to that found in
the Swift-Hohenberg equation. In the chaotic phase shown in Figure
(\ref{fig:dks1}), one finds that even coarse-graining once does not give
us a result satisfactorily close to the benchmark.  This is not 
surprising given that the chaotic phase has positive Liapunov exponent as
discussed before.  In Figure (\ref{fig:dks1}), the benchmark was
obtained using $dx=\pi/16$, $dt=0.0001$ to time $t=16$, whereas the
coarse graining and sampling methods used $dx=\pi/8$, $dt=0.0002$.
Of course, the interesting question to address is whether the statistical
properties of the dynamics are faithfully reproduced by coarse-graining
in a way that is superior to sampling methods.

\section{Discussion}

We have presented a feasible alternative to the numerical solution of
differential equations, in which the continuum limit is attained from a
sequence of coarse-grained function values rather than a sequence of
sampled functions.  The numerical results that we and others\cite{kw,kc}
have obtained are promising and a fully systematic study is clearly
warranted.  In particular, useful acceleration of appropriate
problems is possible without loss of accuracy or even the need to use
adaptive methods.  

There are many issues left unresolved by our work (which is why it has
remained unpublished until this timely occasion).  First and foremost,
what is the correct interpretation to place on the RG scheme when
$\kappa < \infty$?  It can be shown that a literal interpretation
requires a {\it stochastic\/} coarse-graining transformation; in this
case, should one (in principle) average over realisations of the
coarse-graining noise in order to obtain the appropriate solution of the
PDE?  Is it correct to take the mean of this distribution, which seems
to offer a justification for using an exact (i.e. $\kappa=\infty$)
coarse-graining procedure with a $\kappa=2$ formula for the dynamics?
How can one properly extend the philosophy espoused here to nonlinear
problems in a systematic way?

Katz and Wiese\cite{kw} implicitly addressed these issues by deriving the coarse-grained
equations of motion from a postulated action functional $S$, which they varied with
respect to {\it all\/} its arguments. 
In fact their procedure
is equivalent to ours, and their action $S$ can be derived
from our Equation (\ref{eqn:coarse}) by making the Hubbard-Stratonovich transformation
\begin{equation}
e^{-2\kappa W_{\bf n}^2} = \int\frac{dx}{\sqrt{8\pi\kappa}}e^{-x^2/(8\kappa) + ixW_{\bf n}}
\end{equation}
with 
\begin{equation}
W_{\bf n}\equiv U({\bf n},t) -  \frac{1}{a^d}
\int_{-a/2}^{a/2}d^dx\, u({\bf x}+{\bf n}a,t),
\end{equation}
leading to the result that 
\begin{equation}
e^{-\frac{1}{2}\sum_{\bf n} U({\bf n},t) \Delta_a U({\bf n},t)} = \int Du\,D\lambda\, e^{-S\{U({\bf n},t),
u({\bf n},t),\lambda ({\bf n},t)\} }
\end{equation}
where
\begin{eqnarray}
S\{U({\bf n},t),u({\bf n},t),\lambda ({\bf n},t)\} =&& \frac{1}{2}\int d^dx \,\big\{ u(x)\partial_x^2 
u(x) \nonumber\\
&&+  \sum_{\bf n} \left(\frac{\lambda^2}{8\kappa} - i\lambda W\right)\big\}
\end{eqnarray}
is the action (3.5) from reference \cite{kw}.  
Such a functional integral formulation does not possess a small parameter in which to make a loop
expansion based around the classical action 
\begin{equation}
\frac{\delta S}{\delta u} = \frac{\partial S}{\partial \lambda({\bf n},t)} = 0.
\end{equation}
Furthermore, only by requiring the additional constraint
\begin{equation}
\frac{\partial S}{\partial U({\bf n})}=0
\end{equation}
can one use the coarse-grained equation of motion for any $\kappa$ consistently with the constraint
of Equation (\ref{eqn:coarsegrainvar}) (eqn. (3.8) of reference \cite{kw}).

Despite these questions, we are optimistic that the spirit of the
program initiated by Leo P. Kadanoff in critical phenomena will be
equally useful in the fields of pattern formation and partial
differential equations.

\acknowledgements

NG wishes to take this opportunity to thank Leo Kadanoff for providing
encouragement to young scientists and scientific leadership to the
physics community.  The authors thank Leo Kadanoff for bringing
reference \cite{kc} to our attention after the work in this article had
been performed and Joel Lebowitz for inviting us to contribute to this
issue.  This work was supported in part by National Science Foundation
grant NSF-DMR-93-14938 (NG \& QH) and by EPSRC grant GR/K/79307 (AM).

\end{document}